\documentclass[aps,prb,onecolumn,superscriptaddress,notitlepage]{revtex4-1}

\usepackage{amsmath,amssymb,amsfonts,latexsym,fancyhdr}
\usepackage{xcolor,graphicx}
\usepackage{mathtools}

\newcommand{\canc}[1]{}

\begin{document}

\title{Interacting electrons in a flat-band system within the Generalized Kadanoff-Baym Ansatz}


\author{F. Cosco}
\affiliation{Quantum algorithms and software, VTT Technical Research Centre of Finland Ltd, Tietotie 3, 02150 Espoo, Finland}
\author{R. Tuovinen}
\affiliation{Department of Physics, Nanoscience Center
P.O. Box 35, 40014 University of Jyväskylä, Finland}
\author{N. Lo Gullo}
\affiliation{Dipartimento di Fisica, Universit\`a della Calabria, 87036 Arcavacata di Rende (CS), Italy}
\affiliation{INFN, Sezione LNF, gruppo collegato di Cosenza}

\begin{abstract}
This work reports the study of the spectral properties of an open interacting system by solving the Generalized Kadanoff-Baym Ansatz (GKBA) master equation for the single-particle density matrix, namely the time-diagonal lesser Green function. 
To benchmark its validity, the solution obtained within the GKBA is compared with the solution of the Dyson equation at stationarity.  In both approaches, the interaction is treated within the self-consistent second-order Born approximation, whereas the GKBA still retains the retarded propagator calculated at the Hartree-Fock and wide-band limit approximation level. 
The model chosen is that of two leads connected through a central correlated region where particles can interact and utilize the stationary particle current at the boundary of the junction as a probe of the spectral features of the system.
The central region is chosen as the simplest model featuring a degenerate ground state with a flat band. The main result is that the solution of the GKBA master equation captures well the spectral feature of such system and specifically the transition from dispersionless to dispersive behavior of the flat-band as the interaction is increased. Therefore the GBKA solution retains the main spectral features of the self-energy used even when the propagator is at the Hartree-Fock level.
\end{abstract}

\maketitle

\section {Introduction}
Lately, we have been observing a surge in interest and attention towards the out-of-equilibrium dynamics of quantum many-body systems. In this context, the term "out-of-equilibrium" encompasses a broad spectrum of non-trivial dynamical effects, according to the specific field of research or community, triggered by internal effects such as non-linear many-body interactions or external perturbations such as continuous coherent driving \cite{Foieri2010,Sentef2015,Purkayastha2017,Kalthoff2019,Honeychurch2019} or short and strong pulses \cite{Freericks2009,Eckstein2013,Sentef2013,Schueler2016,Mor2017,Perfetto2019,Balaz2023}. Additionally, it extends to non-trivial effects triggered by the coupling to macroscopic systems, acting as thermal reservoirs, which induce particle and energy exchanges \cite{Myohanen2012, Latini2014, Antipov2017, Ridley2018, Ridley2019, Basilewitsch2019,Covito2020, Cohen2020, Dutta2020, Ridley2022}. The field of application of these studies is broad, both theoretically and experimentally, encompassing out-of-equilibrium phases~\cite{Murakami2017,Sentef2017,Li2018,Sentef2018,Topp2018}, pump-probe spectroscopy with a resolution approaching already the femtosecond time-scale
~\cite{Fausti2011,Denny2015,Mitrano2016,Werdehausen2018,  Niedermayr2022, Gupta2023}, band-gap and Floquet engineering~\cite{Lindner2011,Wang2013,Mahmood2016,Huebener2017,Kennes2019,Topp2019}, transport in correlated systems~\cite{Kaiser2014,Hu2014,McIver2020,Talarico2020,Portugal2021}, equilibration and thermalization in strongly correlated materials~\cite{Eckstein2011,Ligges2018,Peronaci2018}, quantum gases \cite{Johnson2016,LoGullo2016,Cosco2017,Cosco2017Pr,Cosco2018Me,Cosco2018AU,Settino2020AU},  relaxation in nano-structures~\cite{Cassette2015,Kemper2018}, and quantum optomechanics \cite{Aspelmeyer2014} and levitodynamics \cite{Cosco2021,Rusconi2022,Ballestero2023}.

Nevertheless, despite the broad spectrum of applications, developing a comprehensive and numerically feasible theoretical framework to describe out-of-equilibrium many-body systems remains challenging. The inherent complexities that contribute to this challenge, such as many-body interactions, external time-dependent fields, and the potential for exchanging energy and matter with the environment, are the same elements that make exploring these dynamics intriguing and a key factor for explaining the wide variety of observed phenomena. Understanding the role of these elements and developing an adequate way to treat them is crucial to give solid ground to new predictions.

In this context, various system-specific approaches have emerged, each presenting its own strengths and limitations. Regardless of the details of the system, each of these methods tends to capture some specific features of the physics of interest, being it many-body correlations, coupling with external baths or dynamically emerging properties in a time-dependent setup.

A special class of numerical methods relies on perturbative approaches to account for external time-dependent fields or many-body interactions. Prominent examples include non-equilibrium Green functions (NEGFs)~\cite{Stefanucci2013}, dynamical mean field theory (DMFT)~\cite{Aoki2014}, and time-dependent density functional theory (TD-DFT)~\cite{Kurth2005}. These techniques operate on a perturbative basis with respect to certain parameters, such as the many-body interaction, coupling to the leads, or tunnelling energy within the system. Notably, these methods facilitate the inclusion of system-environment correlations. Their application remains largely unaffected by the system's geometry and dimensionality, as their computational complexity scales with the space dimension rather than the size of the Hilbert space.

However, the non-equilibrium Green function approach is a computationally demanding method. This is due to the temporal structure structure of the Dyson equation and/or the integro-differential nature of Kadanoff-Baym equations (KBEs). 

In the attempt to develop a NEGF-method with a lower computation cost, but still beyond a semiclassical Boltzmann equation, Ref.~\cite{Lipavsky1986} introduced the so-called Generalized Kadanoff-Baym Ansatz (GKBA). The starting point of this new framework is the KBEs, but by introducing an Ansatz for the retarded propagator it is shown that the two-time structure of the equations can be lifted and it is possible to derive a master equation for the system's density matrix, which has attracted a renewed interest in recent times \cite{Spicka2005, Balzer2013gkba, Hermanns2012, Hermanns2013} thanks to the availability of supercomputers.

Furthermore, improvements to the method have been  proposed constantly from its original formulation~\cite{Haug1996,Bonitz1996,Bonitz1999,Kwong1998,Pal2009,Latini2014,Hopjan2018,Perfetto2018CHEERS}, with the latest works achieving promising progress and tackling one by one all of the inherent limitations of the original GKBA, such as the inclusion of initial correlations~\cite{Karlsson2018,Hopjan2019,Bonitz2019} and the possibility of widening the allowed, in terms of numerical efficiency, many-body perturbation schemes~\cite{Schlunzen2017,Schlunzen2019}. 

Notably, the computational cost of the GKBA approach has been further reduced to a linear in time scaling by mapping the integro-differential master equation to a set of coupled differential equations~\cite{Schlunzen2020, Joost2020, Karlsson2021, Tuovinen2023}.

Due to the simplified structure of the GKBA approach, it is commonly concluded that the respective solution of the equations of motion is incapable of capturing spectral features beyond the Hartree-Fock approximation. In other words, even when the master equation incorporates higher-order corrections, such as those from the second-order Born or third order T-matrix approximations in the collision integrals, the GKBA is thought to fall short in accurately representing the system's spectrum beyond the Hartree-Fock (mean-field) level.

In this study, we argue that the GKBA can effectively capture effects arising from a correlated many-body spectrum when the stationary particle current is employed as a probe for such properties and when the propagator is maintained at the Hartree-Fock level.

To demonstrate this capability of the GKBA, we consider a system undergoing significant changes in spectral properties due to many-body interactions, which can only be accurately described by going beyond the Hartree-Fock self-energy.

 Specifically, we investigate a system whose spectrum undergoes a transition from a flat band to a dispersive band as a result of many-body interactions. Related investigations have recently been carried out in Refs.~\cite{Pyykkonen2021,Pyykkonen2023}. We compare the outcomes of the stationary state of the HF-GKBA master equation with the solution of the stationary state of the full Dyson equation~\cite{Talarico2019}, illustrating that any observed transport features within the GKBA extend beyond the Hartree-Fock spectrum which does not allow for any transport phenomena to occur.


\section {GKBA: Open Interacting Systems}
\label{sec:GKBA}
This section provides an overview of the key characteristics of the solution to the GKBA master equation. For a deeper exploration, readers are directed to the original study~\cite{Lipavsky1986}, as well as a more recent work~\cite{Hermanns2012} which offers a pedagogical derivation and explores applications in inhomogeneous systems. In this context, the derivation seeks to emphasize specific aspects of the GKBA solution that will underpin the subsequent discussion of the primary findings in this work.

For the sake of definiteness, we consider a fermionic system, interacting via a two-body interaction, coupled to fermionic baths. The dynamics of such a system is described by the following general time-dependent Hamiltonian:
\begin{equation}
\hat H(t) = \sum_{i,j}^{} h_{ij}(t) \hat c^\dagger_i \hat c_j
+\frac{1}{2}\sum_{ijkl}u_{ijkl}(t) \hat{c}_i^\dagger \hat{c}_j^\dagger \hat{c}_k \hat{c}_l
+  \sum_{i,k,\alpha} \left[T_{ik}^{\alpha}(t) \hat c^\dagger_i \hat d_{\alpha,k}+ T_{ik}^{\alpha*}(t) \hat d^\dagger_{{\alpha ,k}} \hat c_i\right] + \sum_{\alpha,k} \epsilon_{\alpha,k}  \hat d^\dagger_{\alpha,k} \hat d_{\alpha,k},
\label {eq:tot-ham}
\end{equation}
where $\hat c^\dagger_i \ (\hat c_j)$ are the creation (annihilation) operators of the correlated central system, whereas $\hat d^\dagger_{\alpha, k} \ (\hat d_{\alpha,k})$ are the creation (annihilation) operators of baths, with $\alpha$ labeling them. Furthermore, $h_{ij}(t)$ is the time-dependent single-particle Hamiltonian, $u_{ijkl}(t)$ the two-body interaction tensor, and $T^{\alpha}_{ik}(t)$ is the time-dependent coupling matrix between the modes of the system and the modes of each environment $\alpha$.

Within the NEGF formalism the primary object of interest is the single-particle Green function (SPGF) defined as 
\begin{equation}
\label{eq:spgf}
G_{ij} (z; z') = -\mathrm {i}  \left\langle \mathcal {T}_{c} \hat c_{i} (z) \hat c_{j}^\dagger (z')\right\rangle
\end{equation}
where $z$ and $z'$ are complex variables and $\mathcal T_c$ is the time-ordering operator which orders time over the Schwinger-Keldysh contour.

The Green function satisfies the equation of motion
\begin{equation}
(\mathrm i \partial_z-h(z))G_{} (z;z') =\delta_c (z,z)+ \int \mathrm d\bar z \,  \varSigma (z;\bar z) G (\bar z;z'),
\label{eq:KBE-contour}
\end{equation}
where we have introduced $\varSigma(z;z')$ as the self-energy kernel. In the self-energy we can further distinguish two contributions $\varSigma = \varSigma_{\mathrm {MB}}+\varSigma_{\mathrm {emb}}$.  The first term is the many-body self-energy, accounting for the effects of the many-body interaction between particles within the central region, and the second term accounts for the coupling of the system to the environment and it is generally referred to as embedding self-energy. This latter term is responsible for the exchange of energy and particles between the central region and the external leads.

The equations of motion for the real-time components of the SPGF are derived from Equation~\eqref{eq:KBE-contour} by applying the  Langreth rules. For the retarded and advanced component we obtain
\begin{equation}
(\pm \mathrm i \partial_t-h(t))G^{R/A} (t;t') = \delta(t,t')+ \int \mathrm d\bar t \, \varSigma^{R/A
} (t;\bar t) G^{R/A} (\bar t;t'),
\label {eq:propagator}
\end{equation}
and for the greater and lesser components we get
\begin{align}
(\mathrm i \partial_t-h(t))G^{\lessgtr} (t;t') = I^{\lessgtr}(t,t'), \label{eq:KBE-time1}\\
G^{\lessgtr} (t;t')(-\mathrm i \stackrel{\leftarrow}{\partial}_{t'}-h(t')) = I^{\lessgtr}(t,t').
\label{eq:KBE-time2}
\end{align}
In Eqs.~\eqref{eq:KBE-time1} and~\eqref{eq:KBE-time2} the time-dependent terms on the right side of the equations, i.e. $I^{\lessgtr}(t,t')$, are the  collision integrals and are related to the self-energy and SPGF via the following integral expressions
\begin{align}
 I^{<}( t, t')=\int \mathrm d\bar t \, \left[ \varSigma^{<
} (t;\bar t) G^{A} (\bar t;t') + \varSigma^{R} (t;\bar t) G^{<} (\bar t;t')\right]\\
 I^{>}( t;t')=\int \mathrm d\bar t \, \left[G^{<
} (t;\bar t) \varSigma^{A} (\bar t;t') + G^{R} (t;\bar t) \varSigma^{<} (\bar t;t')\right].
\label{collint}
\end{align}

Collectively, Equation~\eqref{eq:KBE-time1},\eqref{eq:KBE-time2}, and\eqref{eq:propagator} constitute the Kadanoff-Baym equations (KBEs). In the more general framework, these equations extend to the equations of motion for both the right and left components of the Keldysh Green function~\cite{Stefanucci2013}, which are neglected trough our work together with the evolution along the imaginary track. However, solving the KBEs poses a computational challenge, particularly for sizable systems and/or extended durations, owing to the inherent double-time structure of the involved objects.

The GKBA was devised with the aim of simplifying these inherent complexities in the KBEs, thereby mitigating the computational burden required for their resolution~\cite{Lipavsky1986}. In broad terms, the key idea behind this approach involves a decoupling of the time-diagonal components of the SPGF, namely the single-particle density matrix of the system $\rho (t)=- \mathrm i G^{<} (t,t)$, from the time off-diagonal ones and revolves around the solution of the the integro-differential master equation
\begin{equation}
\frac {d} {dt}  {\rho (t)}  + \mathrm {i} \left [h_{HF}(t),\rho (t) \right ] = - \mathrm {i} (I (t) +{\rm h.c.}),
\label{eq:master-equation}
\end{equation}
where we have incorporated mean field effects within the single-particle dynamics trough the Hartree-Fock (HF) Hamiltonian $h_{HF,ij} (t)= h(t)+ \sum_{m,n} w_{imnj} (t) \rho_{nm} (t)$, with $w_{imnj} (t) = 2 u_{imnj}(t)- u_{imjn} (t)$. Analogously to \eqref{collint}, the collision integral is the convolution between the Green function and the self-energy
\begin{equation}
I(t) = \int \mathrm{d} \bar{t} [ \varSigma^>(t,\bar{t}) G^<(\bar{t},t) - \varSigma^<(t,\bar{t}) G^>(\bar{t},t) ].
\end{equation}
On one hand, the collision integral incorporates both many-body effects beyond the mean-field and coupling to the environment. On the other hand, its calculation requires the knowledge and the handling of the greater and lesser Green functions, represented as $G^\lessgtr (t,t')$, at different time times. As it is, solving \eqref{eq:master-equation}, requires solving all the KBEs concurrently. However, the GKBA reduces the computational cost of solving \eqref{eq:master-equation} by decoupling it from, and/or making unnecessary,  solving the equation of motion for the lesser/greater components. This is achieved introducing the following ansatz
\begin{equation}
G^{\lessgtr}(t,t') \approx \mathrm {i} \left [G^{R} (t,t') G^{\lessgtr} (t',t')-G^{\lessgtr} (t,t) G^{A} (t,t')  \right].
\label{eq:lg-GKBA}
\end{equation}
In this framework, the lesser and greater components of the Green functions depend exclusively on the retarded/advanced propagators and the single-particle density matrix, as $G^{>}(t,t)=-\mathrm i ({1}- \rho(t))$. Nonetheless, the GKBA respects the causal structure and the identity $G^>-G^<=G^R-G^A$ still holds.

To move forward, the ansatz needs a suitable approximation for the retarded/advanced propagators. 
For closed systems, the most common choice is to compute the retarded Green function at the Hartree-Fock (HF) level. Hereafter we will refer to the resulting approximation as the HF-GKBA, which has the main advantage of not adding computational complexity to the scheme for solving  the GKBA master equation for the reduced single-particle density matrix. Other possibilities have been studied and put forward~\cite{Bonitz1996,Bonitz1999,Pal2009} which allow to go beyond the HF approximation yet retaining a substantial computational advantage over solving Equation~\eqref{eq:propagator}. 
The HF-GKBA approach has been successfully applied to describe closed many-body systems~\cite{Hermanns2014, Schlunzen2017, Bostrom2018, Perfetto2018, Tuovinen2019pssb, Schueler2019, Perfetto2019, Perfetto2020, Murakami2020, Schueler2020}, as well as open quantum systems~\cite{Latini2014,Hopjan2018,Bostrom2019,Tuovinen2020,Tuovinen2021JCP,Tuovinen2021NJP}.

However, in open systems, in order to avoid solving Equation~\eqref{eq:propagator}, it is possible to resort to the wide band limit approximation (WBLA) for the embedding self-energy~\cite{Latini2014} which is local in time as the HF self-energy.
More recently, it has been shown that it is possible to drop the WBLA
by going beyond the GKBA, namely adding extra terms to Equation~\eqref{eq:lg-GKBA} effectively correcting the time off-diagonal terms of the lesser and greater Green functions~\cite{Kalvova2017,Kalvova2019,Kalvova2023}.

In what follows, and throughout the work, we use the propagator in the HF+WBLA approximation given by
\begin{equation}
G^{R/A}(t,t') =\mp \mathrm {i} \theta [\pm( t-t')]  T e^{ - \mathrm i \int_{t}^{t'} \mathrm d \bar t \, (h_{HF} (\bar t) - \mathrm i \Gamma/2 ) }
\label{G_ret}
\end{equation}
where we have defined the tunneling rate matrix $\Gamma_{ij}= \sum_{\alpha} \Gamma_{ij}^\alpha$, with $\Gamma_{ij}^\alpha=\int d\omega\sum_k\delta(\omega-\epsilon_k)T_{ik}^{\alpha}(T_{jk}^{\alpha})^*$.
This propagator is the formal solution of Equation~\eqref{eq:propagator} 
with $\varSigma(t,t')=(V_{HF}(t)-\mathrm i \Gamma/2)\delta(t-t')$.

The retarded Green function is not the only quantity affected by the presence of the baths; the state of the system itself does depend on them, e.g. the stationary value of the total number of particles in the central system does depend on the chemical potentials of the baths as well as on their temperatures. The effect of the baths on the state and time-correlations are accounted for via the lesser and greater embedding self-energies through the collision integrals. The components of the embedding self-energies are given by:
\begin{equation}\label{eq:sigmaemlssgtr}
\varSigma_{\mathrm {emb},ij}^{\lessgtr}  (t,\bar t) = \pm  \mathrm {i} \sum_{\alpha} \Gamma_{ij}^\alpha  \int  \mathrm {d} \omega \, f(\beta_\alpha , \pm \mu_\alpha,  \pm \omega)e^{-\mathrm i \omega(t-\bar t)},
\end{equation}
where $f(\beta_\alpha , \mu_\alpha , \omega)$ is the Fermi-Dirac distribution of the the environment $\alpha$, and depends on the inverse temperature $\beta_\alpha $ and chemical potential $\mu_\alpha$ of each bath. In Eq.~\eqref{eq:sigmaemlssgtr} and in the following we consider time-independent electro-chemical potentials; a time-dependent modulation would add an additional phase factor~\cite{Tuovinen2020}.

In the GKBA, for the many-body part of the system Hamiltonian, the HF propagator includes the effect of the interaction at the mean-field level in the time-diagonal component, whereas the collision integral accounts for inclusion of correlations. In this work, we employ the second order Born approximation (2B), for which the lesser and greater self-energies are~\cite{Hermanns2012,Latini2014,Tuovinen2019-2B,Settino2020AU}:
\begin{equation}
\varSigma_{\mathrm {MB},ij}^{2B,\lessgtr}  (t,\bar t)= \sum_{mnpqrs} u_{irpn} (t)w_{mqsj} (\bar t) G^{\lessgtr}_{nm}(t,\bar t) G^{\lessgtr}_{pq}(t,\bar t) G^{\gtrless}_{sr}(\bar t, t).
\label{se-2b}
\end{equation}
The self-energies in Eq. \eqref {se-2b} are advantageously calculated within the GKBA framework, as they do not require time integrals.
In addition to the 2B approximation, in some numerical simulations presented later on we also consider the $T$-matrix approximation (particle-particle channel)
\begin{equation}
\Sigma_{\mathrm{MB},ij}^{T,\lessgtr}(t,\bar{t}) = \mathrm{i}\sum_{kl}T_{ikjl}^\lessgtr(t,\bar{t})G_{lk}^\lessgtr(\bar{t},t),
\end{equation}
where $T$ is constructed via a consideration of the two-particle Green function and the associated Bethe-Salpeter equation~\cite{Schlunzen2019, Joost2020, Pavlyukh2022}.
Under the assumptions made, wide-band-limit for the environmental degrees of freedom, and within the 2B approximation the GKBA master equation retains a computational cost scaling as $\sim N_t^2$, with $N_t$ the number of steps as opposite to the $N_t^3$ scaling of the KBEs.
The approach is therefore very promising as it allows to explore the long-time dynamics of large systems retaining at the same time some of the most appealing features of the NEGFs. 

\section{Transport spectroscopy within the GKBA approach}
\label{sec:curr}


It is compelling to conclude that the GKBA master equation fails in capturing the spectral features of correlated many-body systems and this issue seems conceptually difficult to be overcome without altering the benefits of the approach. Any attempt to include correlations in the retarded Green function would require the solution of Eq.~\eqref{eq:propagator}, possibly ruining the computational advantage gained with the use of the GKBA over the full KBE. Formally, this makes it impossible to include correlation terms in the single-particle spectrum, as is clear from the spectral function 
\begin{equation}
A(\omega)=\underset{T\rightarrow \infty}{\lim}\int d\tau\; A(T+\tau/2,T-\tau/2),
\label{eq:spfunc}
\end{equation}
where $T=(t+t')/2$, $\tau=t-t'$, and $A(t,t')=G^R(t;t')-G^A(t;t')$, where the propagators are calculated at the HF level.  Including higher-order effects in the many-body self-energy appearing in the collision integral, while maintaining the propagator at the HF level could be understood as irrelevant \cite{Hopjan2019} since it will not alter the properties of the spectral function.
However, we argue that the spectral properties of a system can nonetheless appear in and be inferred from  physically relevant quantities other than the explicit spectral function.

In order to explore the spectral properties of the solution of the HF-GKBA master equation, we follow the approach of transport spectroscopy~\cite{Bruus_Flensberg} to get an insight into the spectral properties of the physical system under study. We compare the results obtained with the HF-GKBA with those obtaining by solving the full Dyson equation at stationarity.

The concept behind transport spectroscopy is akin to angle-resolved photoemission spectroscopy (ARPES), where occupied energy levels are examined through photoemission. In this case, however, the focus is on probing the transmission function of the system, closely linked to the spectral function, by scanning across the chemical potentials of the centers of the lead energy band, known as the bias window.


In the NEGFs framework, the particle current for each lead $\alpha$ is given in time domain as~\cite{Meir1992, Jauho1994, Tuovinen2013}:
\begin{equation}
\mathcal{I}_\alpha(t)=2\textrm{Re}\left\{\int d\bar t\; \textrm{Tr}\left[\varSigma_\alpha^<(t,\bar t)G^A(\bar t;t)+\varSigma_\alpha^R(t,\bar t)G^<(\bar t;t)\right]\right\},
\label{eq:curr-time}
\end{equation}
and in frequency domain as:
\begin{equation}
	\mathcal{I}_\alpha=i\int \frac{d\omega}{2\pi}\; \textrm{Tr}\left[\varSigma_\alpha^<(\omega)A(\omega)-\Gamma_\alpha(\omega) G^<(\omega)\right],
\label{eq:curr}
\end{equation}
when we take the long-time limit, and where $\Gamma_\alpha(\omega)=i\left(\varSigma_\alpha^R(\omega)- \varSigma_\alpha^A(\omega)\right)$ and $ A(\omega)=i\left( G^R(\omega)- G^A(\omega)\right)$. While Eq. \eqref{eq:curr} depends explicitly on the spectral function $A(\omega)$ and the Fourier transform of the lesser Green function $ G^<(\omega)$, the expression can be further simplified and rewritten in the more common Landauer-B\"uttiker formula ~\cite{Landauer1957, Buttiker1986}
\begin{equation}
\mathcal{I}_\alpha=\sum\limits_{\beta}\int \frac{d\omega}{2\pi}\; T_{\alpha \beta}(\omega)(f_\alpha(\omega)-f_\beta(\omega)),
\label{eq:LB}
\end{equation}
with the transmission coefficient defined as $T_{\alpha \beta}(\omega)=\textrm{Tr}\left[\Gamma_\alpha(\omega) G^R(\omega)\Gamma_\beta(\omega) G^A(\omega)\right]$ and  the shorthand notation $f_\alpha(\omega)\equiv f(\omega,\beta_\alpha,\mu_\alpha)$ for the Fermi-Dirac distribution of the $\alpha$-th bath.

Although this form has a more intuitive and immediate physical meaning, it reinforces the idea that the GKBA approach is unable to capture spectral features beyond the HF+WBLA due to the fact that the transmission coefficient is exclusively a function of the retarded and advanced Green functions.

However, the derivation of Equation~\eqref{eq:LB} relies on the equality $G^<(\omega)=G^R(\omega)\varSigma^<(\omega)G^A(\omega)$~\cite{Jauho_Haug} 
which is a direct consequence of the solution of Equation~\eqref{eq:propagator} with the full retarded self-energy in the stationary limit.
This expression holds in the absence of bound states, but most importantly it uses the fact that the $G^R(t,t')$ is the solution of Eq.~\eqref{eq:propagator}.

\section {Correlation activated transport in a flat band}
\begin{figure}[t!]
\centering
	\includegraphics[width=\linewidth]{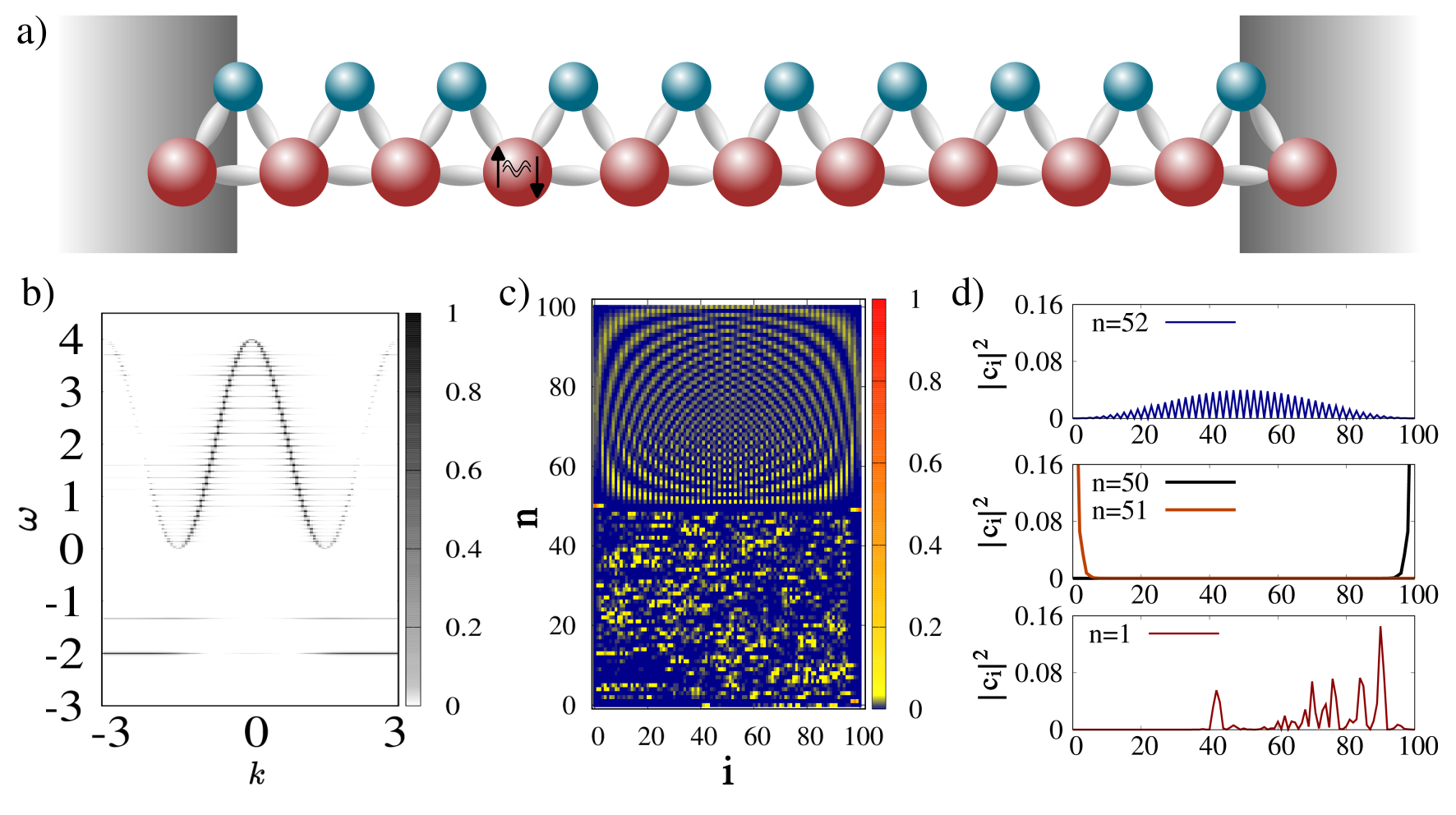}
	\caption{(Color online). The system. Panel {\bf a)}: a schematic representation of the system described by Equation~\eqref{eq:ham-flat} with the central region hosting electrons with spin up and down which interact locally. The two leads on the left and right are modeled as non-interacting Fermi gases with electrons with both spins. The hopping between different sites are schematically represented by the grey areas between the sites and, if the geometry is flattened in one-dimension, the system features nearest-neighbour hopping for all sites, and next-nearest-neighbour hopping between odd sites. Panel {\bf b)} show the spectral function for a non-interacting ($U=0$) and decoupled ($T^\alpha_{ik}$=0) system with $N=101$. There it is visible the highly degenerate ground state, the dispersive band and the in-gap localized states. Panel {\bf c)} shows a density map of the eigenstates corresponding to the system in panel {\bf b)} and panel {\bf d)} shows (from top to bottom) the lowest energy state of the dispersive band, the two in-gap localized states, and two states belonging to the manifold of the ground states.}
	\label{fig:system}
\end{figure}

In this section we introduce the physical model to highlight the properties of the solution of the HF-WBLA-GKBA solution.
The model -- a quasi one-dimensional sawtooth lattice (see Figure~\ref{fig:system}) -- is the simplest of a class of models used to explain the emergence of ferromagnetism in the ground state of itinerant-electrons systems~\cite{Mielke1991,Tasaki1992,Mielke1993} and its physical properties are described by the following Hamiltonian:

\begin{eqnarray}
&\hat H =&\hat H_w +\sum\limits_{\alpha=L,R}\hat H_\alpha+\hat V_\alpha\\
\label {eq:ham-flat}
&\hat H_w =& \sqrt{2}\sum_{i=1,\sigma=\uparrow\downarrow}^{N} \left(\hat c^\dagger_{i \sigma } \hat c_{(i+1) \sigma}+\text{h.c.} \right)+\sum_{i=1,\sigma=\uparrow\downarrow}^{{\lfloor \frac{N}{2}\rfloor}}\left(\hat c^\dagger_{2i \sigma } \hat c_{(2i+2) \sigma } +\text{h.c}\right) -\frac{U}{2}\sum_{i=1,\sigma=\uparrow\downarrow}^{N} \hat c^\dagger_{i \sigma } \hat c_{i \sigma} \nonumber \\
&& +U \sum_{i=1}^{N} \hat c^\dagger_{i \uparrow  } \hat c^\dagger_{i \downarrow} \hat c_{i \downarrow  } \hat c_{i \uparrow }\\
&\hat H_\alpha =& \sum_{k, \sigma=\uparrow\downarrow} \epsilon_{\alpha,k}  \hat d^\dagger_{\alpha, k \sigma} \hat d_{\alpha, k\sigma} \\ 
&\hat V_\alpha =& \sum_{i=1,k, \sigma=\uparrow\downarrow} \left[T_{ik}^{\alpha} \hat c^\dagger_{i \sigma } \hat d_{\alpha,k \sigma}+ T_{ik}^{\alpha*} \hat d^\dagger_{{\alpha, k \sigma }} \hat c_{i \sigma }\right],
\end{eqnarray}
where $\hat c^\dagger_{i\sigma} \ (\hat c_{i\sigma})$ are the creation (annihilation) operators for electrons in the basis labeled by the site $i$ and spin component $\sigma$, with $N$ denoting the total number of sites in the system. The operators $\hat d^\dagger_{\alpha, k \sigma} \ (\hat d_{\alpha,k \sigma})$ are the creation (annihilation) operators of the two different leads, denoted as $L$(eft) and $R$(ight), and labeled by $\alpha$. The first two terms in $\hat{H}_w$ are `diagonal' and `longitudinal' hoppings of strength $\sqrt{2}t$ and $t\equiv 1$, respectively, and the third term accounts for the Hartree shift due to the the two-body interaction $U$ between spin-up and spin-down particles on the same site. The forth term in $\hat{H}_w$ is the on-site Hubbard interaction of strength $U$. The leads are non-interacting, whose single-particle structure is specified by the energy dispersion $\epsilon_{\alpha,k}$ in $\hat{H}_\alpha$. In $\hat{V}_\alpha$, the coefficients $T_{ik}^{\alpha}$ are the coupling energies between the leads $\alpha$ and the system.
In the following, we assume that the left and right leads are coupled to the first-two and last-two sites of the system respectively and described within the WBLA, i.e. $T^L_{ik}= T_L (\delta_{i1}+\delta_{i2})$ and $T^R_{ik}= T_R (\delta_{iN-1}+\delta_{iN})$. 
Furthermore, the two leads are kept in a thermal state at the same temperature, i.e. the inverse temperatures $\beta_L=\beta_R=\beta=10^{3}$, but in general at different chemical potential $\mu_L\neq \mu_R$.
A pictorial representation of the system is shown in Figure~\ref{fig:system}~{\bf a)}. 

For the purpose of our study, it is important to mention that in the non-interacting ($U=0$) and closed-system ($T^\alpha_{ik}$=0) limits, the system features a highly degenerate single-particle ground state (in the thermodynamic limit; $N\rightarrow \infty$) which is separated in energy from a dispersive band by an energy gap of $4t$.
Moreover there are exactly two exponentially localized states with energies within the energy gap.
This can be clearly seen from the spectral function of the system $A(k,\omega)$ shown in Figure~\ref{fig:system}~{\bf b)} where the dispersive and flat bands are clearly visible together with the exponentially localized in-gap states.
A density map of the eigenstate's wavefunctions for the whole spectrum of a system with $N=101$ and open boundary conditions is reported in Figure~\ref{fig:system}~{\bf c)}. From it, one can appreciate the markedly different structure of the eigenstates belonging to the dispersive and flat bands.
Specifically, eigenstates belonging to the dispersive band are delocalized as shown in the top panel in Figure~\ref{fig:system}~{\bf d)}. Instead, states belonging to the highly degenerate manifold of the flat band are cluster localized as shown in the bottom panel of Figure~\ref{fig:system}~{\bf d)}.
The in-gap states are exponentially localized at the edges of the system as shown in the middle panel of Figure~\ref{fig:system}~{\bf d)}.

We will focus on the transport signatures involving only the flat band states by pinning the chemical potentials around these energies. While also the dispersive band and the in-gap localized states are both present in the model, they are thus effectively excluded from the transport calculations.
The reason for this is that, for the purpose of this work, the dispersive band and the delocalized states do not present any difference with respect to previous studies where a comparisons between the GKBA solution and the KBE (or Dyson at stationarity) have been carried out~\cite{Hermanns2014}.
The in-gap localized states give rise to persistent oscillations during the dynamics if they are initially populated, this effect is of no interest for the specific purpose of this work and thus we consider cases in which they are not populated, neither in the initial state nor during the dynamics. This is achieved by controlling the initial chemical potential of the system and the chemical potential of the leads.

To probe the spectral properties of the states in the flat band we simulate a transport spectroscopy experiment~\cite{Bruus_Flensberg}. The key idea is to scan the energy spectrum of the central system by sweeping the applied bias voltage across the energy region of interest. In our simulations this is accomplished by setting the chemical potential of the leads to $\mu_L=\mu+V_b/2$ and $\mu_R=\mu-V_b/2$ where $V_b>0$ is the bias voltage applied across the system.
Unless otherwise stated the bias-window $V_b$ will be kept fixed and the sweeping will be performed by tuning the center of the bias window $\mu$.
For all simulations we chose $V_b$ to be much smaller than the energy separation between the in-gap states and the flat band states; in the case of interacting electrons ($U\neq 0$) we carefully checked that this is true when the Hartree shift is included.

\subsection{Transport within the HF approximation}
Due to the cluster-localized nature of states belonging to the flat band it is worth investigating whether they conduct. In the case of non-interacting fermions we find (not shown) that at stationarity there is no conduction for any value of the center of the bias window $\mu$ around the flat-band energy. This is true for any value of the coupling to the leads and for any bias window width $V_b$, provided it is not larger than half the energy gap separating the flat-band from the higher energy dispersive one.
In the the setup studied this is easily understandable resorting to the transmission function $T(\omega)$ which vanishes for all $\omega$ due to the lack of overlap between states which have a non-vanishing connection with both leads which spatially separated and connected only through the central region.
Therefore, although these states are not exponentially localized, the clustering of the probability density results in a insulating behavior. Also, for electrons occupying a flat energy band, the group velocity vanishes, $d\epsilon_{k}/d_k = 0$, resulting in zero mobility.

\begin{figure}[t!]
\centering
	\includegraphics[width=0.5\linewidth]{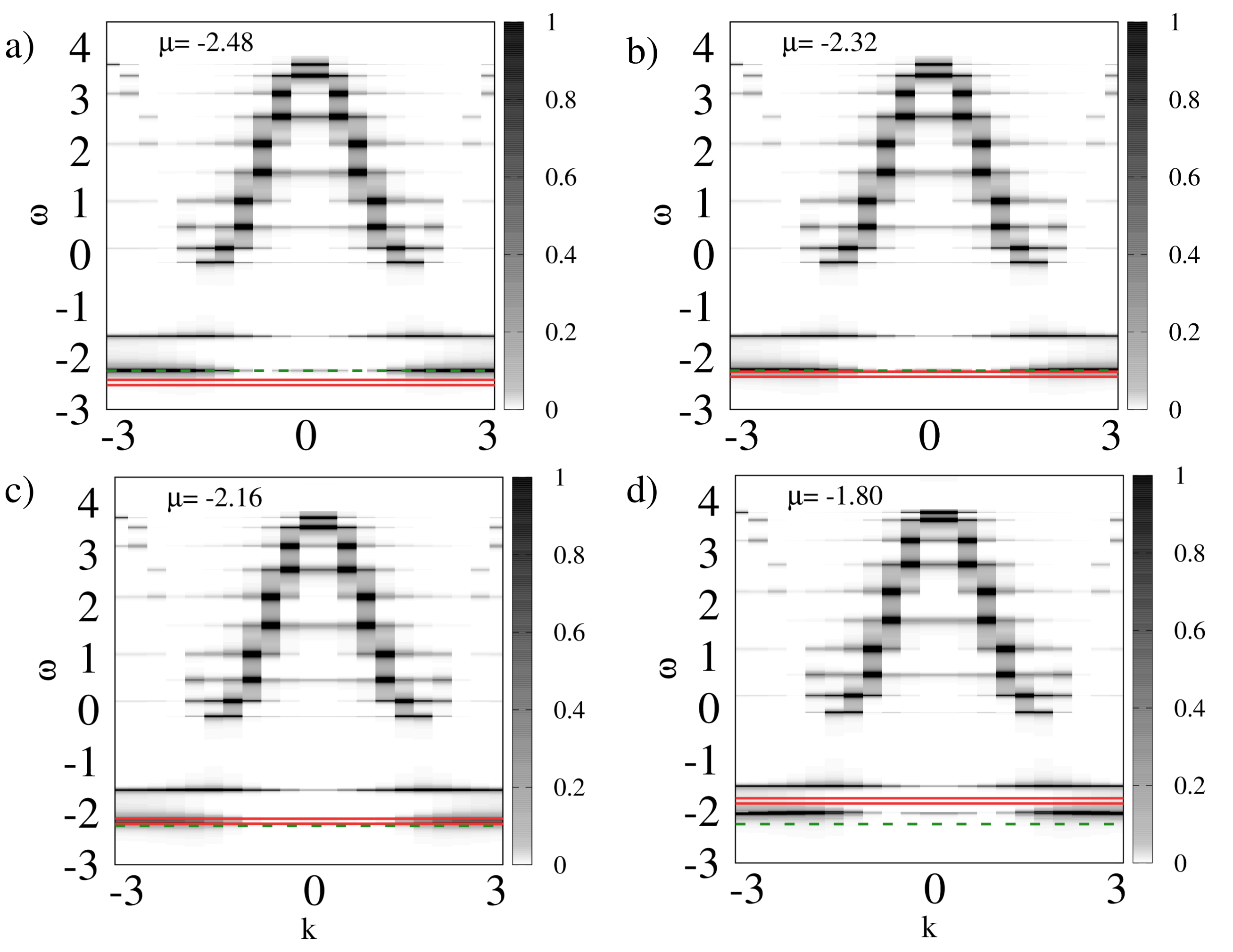}
	\caption{(Color online). HF Spectral Function. Spectral function for a system with $N=21$, $\Gamma=0.1$, $U=0.5$ and $V_b=0.1$ and for different $\mu$. The dashed green line marks the position of the flat band for a non-interacting ($U=0$) and non-coupled $\Gamma=0$ system, whereas the solid red lines mark the position of the left(right) lead's chemical potential $\mu_L=\mu+V_b/2$($\mu_R=\mu-V_b/2$).}
	\label{fig:spec-funcs_HF}
\end{figure}

We found that correlations induced by the many-body interaction restores the conduction. Let us first consider the case in which the many-body interaction in Equation~\eqref{eq:ham-flat} is accounted for within the Hartree-Fock approximation. Because there is perfect symmetry between up,down spin components the Fock component vanishes identically and we are left only with the Hartree term.
The effect of this term, although intuitive, deserves some attention especially in view of our discussion of the dynamics within the HF-WBLA-GKBA master equation in the following section.
To show the effect of such a term, we looked at the spectral function of the system for different $\mu$.
The results are reported in Figure~\ref{fig:spec-funcs_HF} for a system with $N=21$, $\Gamma=0.1$, $U=0.5$ and $V_b=0.1$.
The numerical simulations have been done solving self-consistently the Dyson equation in the frequency domain using a frequency interval $\omega\in [-2\pi,2\pi]$ sampled with $N_\omega=2 \times 10^5$ points.
It is important to mention that in order to achieve convergence, both for the HF and for the second Born self-energies (see next section), we had to use the Pulay mixing as described in~\cite{Rubio2008,Talarico2020}.
Specifically we used a memory of ten previous solutions with a mixing parameter varying between $0.2$ and $0.9$ depending on the conduction properties of the system. Low mixing values, favoring input solutions, were used in the case of non-conductive bands, whereas higher values were needed when the conduction was restored due to interactions. In the case of HF a constant value of 0.5 was used and worked for all explored parameters.

The green dashed horizontal line marks the energy of the flat-band in the non-interacting case to be used as a reference for the interacting one. The two red solid horizontal lines mark the chemical potential of the left and right lead respectively.
In the case in which the average chemical potential of the leads $\mu$ is well below the non-interacting flat-band energy, Figure~\ref{fig:spec-funcs_HF} {\bf a)} there is no shift of the flat-band. This is due to the fact that the system is empty and therefore the interaction is effectively vanishing. As $\mu$ is increased, and the system starts to be filled with electrons, the interaction starts playing a role. Therefore there exists an intermediate interval of values of $\mu$ for which the system gets filled very slowly as $\mu$ is increased and the final position of the flat-band is around its non-interacting value. This is the case in Figure~\ref{fig:spec-funcs_HF} {\bf b} and {\bf c}.
Once $\mu$ becomes large enough as in Figure~\ref{fig:spec-funcs_HF}{\bf d} the position of the flat-band stabilizes and it changes proportionally with the variation of $\mu$.

Despite the above mentioned effect on the spectrum of the system, interactions within the HF approximation do not affect the conduction properties of the system. Regardless of the regime we explored, we found (not shown) that around the flat-band the system still behaves as an insulator and no conduction is observed.
This is primarily due to the fact that the single-particle eigenstates of the flat-band manifold, in this case being the eigenstates of the HF single particle Hamiltonian, retain the same structure as those of the non-interacting system. Specifically they are similar to those shown in Figure~\ref{fig:system} {\bf c)} and {\bf d)}.

\subsection{Transport beyond the HF approximation}
To investigate the effect of many-body correlations on the transport properties of the states in the flat-band we go beyond the HF approximation and include second-Born (2B) diagrams to the self-energy. 
In the case of the Dyson equation this amounts to adding a self-energy to both the equation for the Retarded and Lesser/Greater components of the SPGF, which is added to the embedding self-energies of the leads.
The main effect of adding the 2B self-energy, and therefore many-body correlations, is to restore transport when the center of the bias window is swept across the non-interacting single-particle flat band.

\begin{figure}[t!]
\centering
	\includegraphics[width=0.5\linewidth]{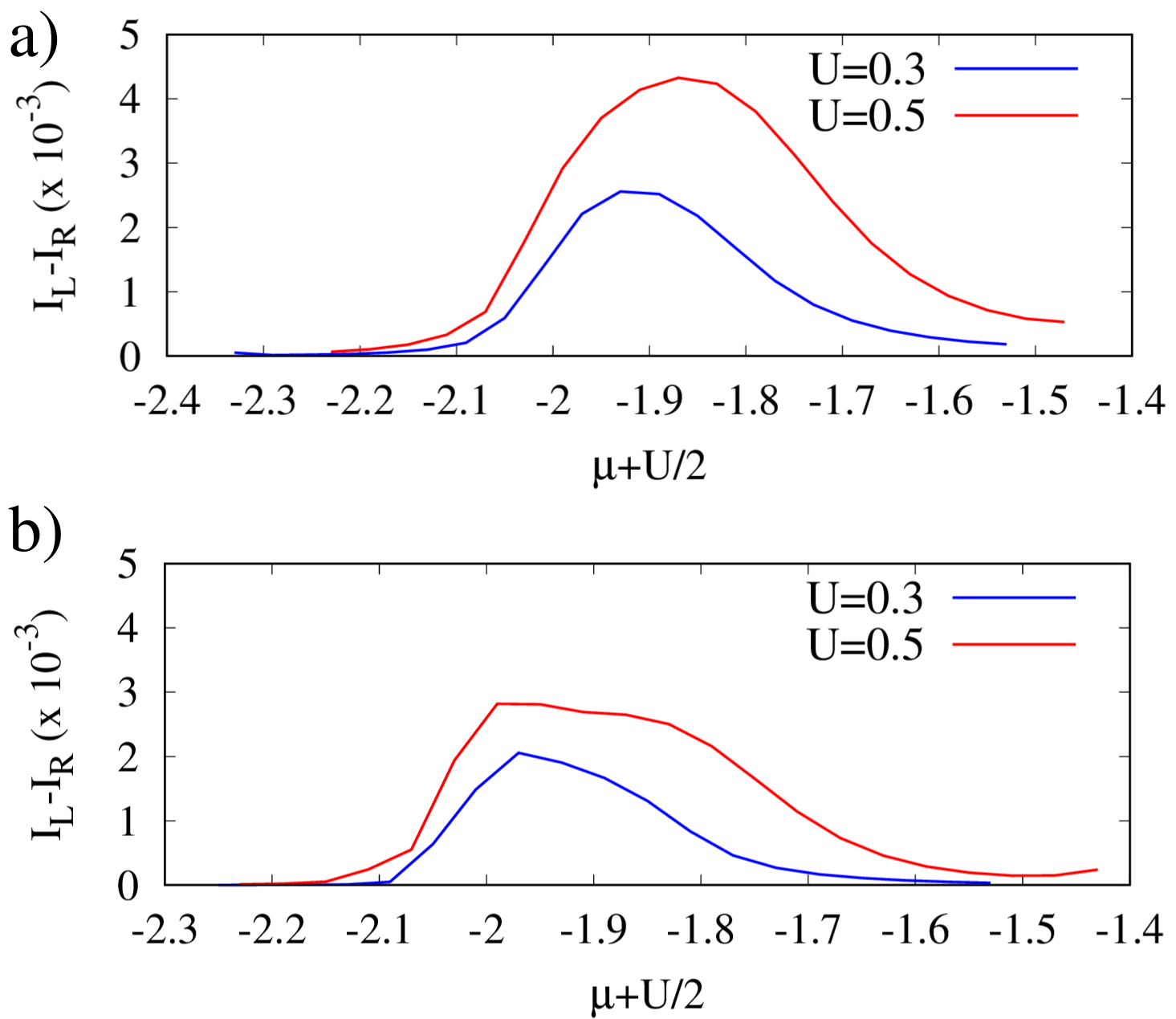}
	\caption{(Color online). 2B particle current. The stationary particle current in the 2B approximation obtained from the solution of the Dyson equation for different interaction strengths, and for $\Gamma=0.1$ (panel {\bf a)}) and $\Gamma=0.5$ (panel {\bf b)})}
	\label{fig:2B_current}
\end{figure}

The interaction here plays a key role in activating the transport of the non-interacting flat-band states. The larger the interaction the more pronounced the transport of particles across the system. This is shown in Figure~\ref{fig:2B_current} where we compare the current across the system as a function of the center of the bias window for different interaction strengths. In particular, the stronger the interaction, the larger the total current in the system.
This behavior is true for both weak and moderately strong couplings to the leads, as shown in the panels {\bf a)} ($\Gamma=0.1$) and {\bf b)} ($\Gamma=0.5$) of Figure~\ref{fig:2B_current}.
For a normal metal, namely in the case of the chemical potential of the system within a dispersive band corresponding to delocalized states, the introduction of a repulsive many-body interaction would result in a suppression of conduction due to inter-particle collisions, especially in low-dimensional systems.
Here instead we observe the activation of transport. This phenomenon is similar to the one observed in aperiodic potentials~\cite{Bordia2017,Settino2020AU} which share with the system considered here the structure of the eigenstates, namely they are cluster localized, as shown in the bottom figure of Figure~\ref{fig:system} panel {\bf d)}.
In aperiodic potentials, and specifically in the case of the Aubry-André model, a moderate interaction results in both an anomalously slow dynamics as well as a restoring of the conduction at the critical point where the system would otherwise be insulating.


\section{Transport within the HF-WBLA-GKBA}
In the previous section we have seen that adding many-body correlations to the description of the system, namely going beyond the HF approximation, results in the restoring of the transport properties of the system at those energies. From the theoretical point of view, and using the Meir-Wingreen formula for the current, it is possible to infer that this behavior is due to the appearance of single-particle energy states which participate in the conduction. The link between the transport properties of a system and its single-particle energy spectrum can be exploited to gather information about the latter; this is the underlying principle of transport spectroscopy~\cite{Bruus_Flensberg}. In this section we will look at the current through the system using the solution of the HF-WBLA-GKBA.

The simulations with the HF-WBLA-GKBA master equation have been done by evolving the system starting with an initial state which has no correlation between the central region and the leads, and the system is in the ground state of the non-interacting Hamiltonian. It follows an initial evolution in which both the interactions and the coupling to the leads is slowly switched on. The switching period, in units of inverse hopping $t^{-1}$ is $10^4$. The lead chemical potentials, together with a bias window $V_b$ are constant and nonzero throughout the time evolution. After the switching period, the system is simply evolved until it reaches a non-equilibrium stationary state.
The latter is checked by looking at the sum of the currents in the left and right lead, and in particularly we have considered the system to have reached its stationary state if the sum of the currents is smaller than $10^{-7}$.
The total evolution time, excluding the initial preparation, was typically of the order of $9\times 10^4$ for all simulations we have performed.
This is somehow expected due to the almost-insulating nature of the flatband states, therefore saturation towards a stationary state requires long times, which typically increase with the size of the system.
\begin{figure}[t!]
\centering
	\includegraphics[width=0.9\linewidth]{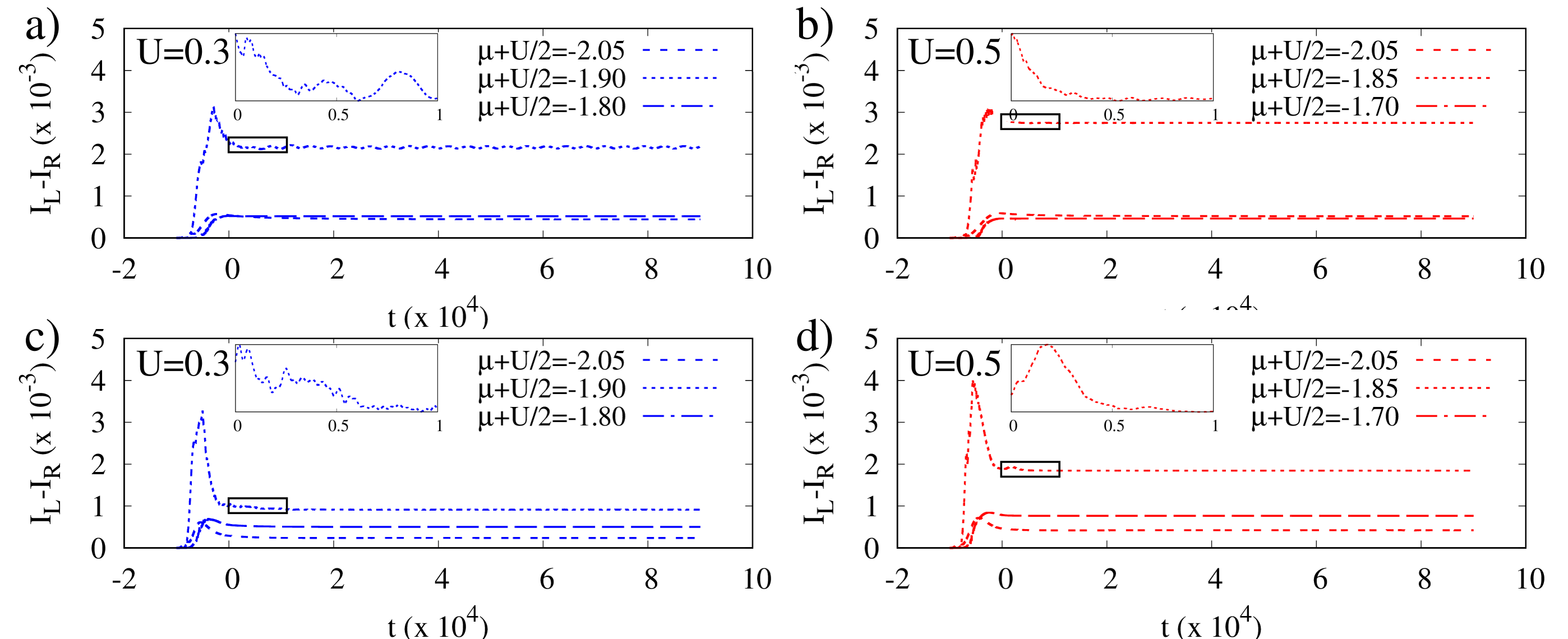}
	\caption{(Color online). Time dependent currents. The evolution of the current for $\Gamma=0.1$  (panel {\bf a)} and {\bf b)}) and $\Gamma=0.5$ (panel {\bf c)} and {\bf d)}) and different interactions $U$. In each panel it is shown the behavior of the current for different values of the chemical potential $\mu$ (see text). The insets show a zoom in of the relaxation in the time interval $t\in[0,10^4]$ marked by the rectangle on the different curves.}
	\label{fig:time_currs_GKBA}
\end{figure}

Another aspect which is important to mention is that in some of the GKBA simulations we performed with the second Born approximation we observed numerical instabilities. Similar violations (with other many-body approximations) of physical properties, e.g., the positivity of occupation numbers have recently been reported~\cite{Pavlyukh2022,Joost2022}, and this is still an open problem. For the purpose of the present study, we have performed GKBA simulations also with the $T$-matrix approximation in the particle-particle channel, for which the lowest-order diagram is the same as the second Born one. (Due to the local Hubbard interaction in our model, the exchange contributions vanish.) In the $T$-matrix case we did not encounter any numerical issues for this model. Our aim here is to assess whether the GKBA approach can produce qualitatively similar result of interaction-induced transport concerning the flatband states, independent of how accurate the many-body approximation is when compared against the exact solution.

In Figure~\ref{fig:time_currs_GKBA} we show a selection of the time dependent currents obtained from the HF-WBLA-GKBA solution with the $T$-matrix approximation. The initial switching of the coupling to the leads and of the interaction is also shown, the relaxation starts at $t=0$ in all figures, whereas negative times $t\in[-10^4,0)$ correspond to the switching on of the couplings and many-body interaction.
The first thing we notice is that relaxation occurs on a much longer time scale of the one set by the relaxation time $\tau\approx \Gamma^{-1}$, which in our case is of the order of ten in units of $t^{-1}$. Instead for all the parameter ranges we explored, we consistently found that relaxation towards the steady state requires a time which is at least three order of magnitude larger than the relaxation time defined above.
As anticipated, this is due to the localized nature of the states belonging to the flatband manifold. This feature is shared with the non-interacting ($U=0$) case and with the case in which interactions are treated with the Hartree-Fock approximation.
After the transient we observed that the system reaches a non-equilibrium steady state characterized by a non-vanishing steady current flowing through the central region.
This is an important result for our study: in the case of the Hartree-Fock approximation the steady state current vanishes independently of the coupling strength to the reservoirs $\Gamma$ and of the many-body interaction $U$.
The fact that in the $T$-matrix approximation the steady state currents is non-vanishing implies that the HF-WBLA-GKBA approximation is able to capture important spectral features arising from the many-body correlations.

\begin{figure}[t!]
\centering
	\includegraphics[width=0.5\linewidth]{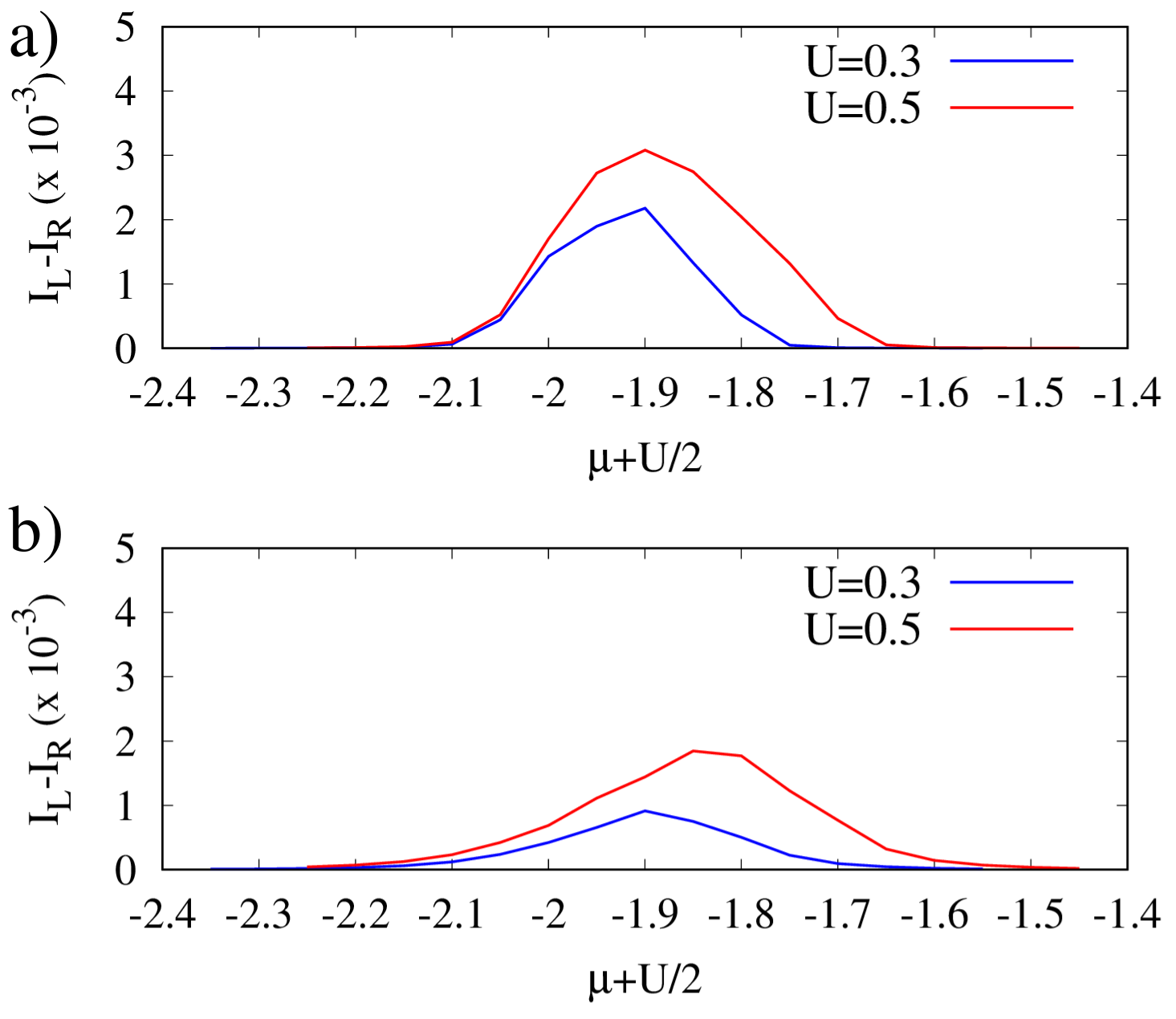}
	\caption{(Color online). $T$-matrix particle current. The stationary particle current in the $T$-matrix approximation obtained from the solution of the HF-WBLA-GKBA for different interaction strengths, and for $\Gamma=0.1$ (panel {\bf a)}) and $\Gamma=0.5$ (panel {\bf b)})}
	\label{fig:currs_GKBA}
\end{figure}

In Figure~\ref{fig:currs_GKBA} we show the steady state current for different values of the interaction as a function of the center of the bias window $\mu$. The top (bottom) panel refer to $\Gamma=0.1$ ($\Gamma=0.5$).
We immediately notice the similarity with the behavior obtained with the solution of the Dyson equation shown in Figure~\ref{fig:2B_current}.
Specifically we notice that for both coupling strengths $\Gamma$ the current increases with increasing $U$. This confirms that indeed the HF-WBLA-GKBA captures the change in the spectral properties of the system.
Another interesting feature which is captured is the reduction of the current with increasing $\Gamma$. This might be due by the fact that a larger $\Gamma$ results in a larger broadening of the density of states of the system, therefore reducing the magnitude of the integrated density of states inside the bias window of the leads~\cite{Thygesen2008,Tuovinen2021JCP}. 
This is in fact yet another hallmark of the fact that the HF-WBL-GKBA, can capture spectral features which are beyond the HF ones and which come from the inclusion of the many-body correlations through the Lesser and Greater components in the collision integrals.

\section{Conclusions}

In this work, we compared the outcomes of the Generalized Kadanoff-Baym Ansatz (GKBA) master equation with the stationary solution derived from the full two-times Dyson equation. Specifically, we studied a transport setup featuring two leads, approximated in the wide band limit, interconnected through a central region modeled as a quasi one-dimensional sawtooth lattice which is an effective model introduced to explain the emergence of ferromagnetism in the ground state of itinerant-electrons system.

By using the stationary current as figure of merit, our analysis revealed that the GKBA master equation, computed with a Hartree-Fock propagator and higher order self-energy in the collision integral, captures  physical effects due to spectral features absent in the mean field spectral function. 
To support our argument we highlighted how in the chosen model transport is driven exclusively by many-body effects beyond the mean-field approximation. In addition, we observed the transport signatures related to the flat-band states provide a rich platform for non-equilibrium many-body effects. We will address this topic more thoroughly in a forthcoming paper.

In simpler terms, our results provide numerical evidence that the GKBA succeeds in capturing spectral features beyond those included in the Hartree-Fock propagator, through the higher order self-energies included in the collision integral.
This finding establishes the GKBA as a valuable instrument for simulating and describing ARPES and pump-probe experiments, where the crucial quantity for signal reconstruction corresponds precisely to the lesser component of the Green function. Furthermore, although our focus primarily rested on the stationary state, our methodology seamlessly extends to investigate time-resolved transport in general correlated quantum systems. This extension holds significance in addressing transiently emerging phenomena, such as superconductivity and Majorana physics~\cite{Stefanucci2010, Jiang2011, Weston2015, Francica2016, Tuovinen2016PNGF, Thakurathi2017, Dehghani2017, Claassen2019, Tuovinen2019NJP}.

Our work, coupled with recent advancements and improvements of the method, highlight the role of the GKBA as a potent and reliable tool for studying out-of-equilibrium phenomena in both many-body open and closed quantum systems.


\medskip
\textbf{Acknowledgements} \par 
Numerical simulations were performed using the Finnish CSC facilities under the Project no. 2000962 (``Thermoelectric effects in nanoscale devices'').



\medskip

\begin{thebibliography}{100}
\providecommand{\url}[1]{\texttt{#1}}
\providecommand{\urlprefix}{URL }

\bibitem{Foieri2010}
F.~Foieri, L.~Arrachea,
\newblock \emph{Phys. Rev. B} \textbf{2010}, \emph{82} 125434.

\bibitem{Sentef2015}
M.~A. Sentef, M.~Claassen, A.~F. Kemper, B.~Moritz, T.~Oka, J.~K. Freericks,
  T.~P. Devereaux,
\newblock \emph{Nature Communications} \textbf{2015}, \emph{6}, 1 7047.

\bibitem{Purkayastha2017}
A.~Purkayastha, Y.~Dubi,
\newblock \emph{Phys. Rev. B} \textbf{2017}, \emph{96} 085425.

\bibitem{Kalthoff2019}
M.~H. Kalthoff, D.~M. Kennes, M.~A. Sentef,
\newblock \emph{Phys. Rev. B} \textbf{2019}, \emph{100} 165125.

\bibitem{Honeychurch2019}
T.~D. Honeychurch, D.~S. Kosov,
\newblock \emph{Phys. Rev. B} \textbf{2019}, \emph{100} 245423.

\bibitem{Freericks2009}
J.~K. Freericks, H.~R. Krishnamurthy, T.~Pruschke,
\newblock \emph{Phys. Rev. Lett.} \textbf{2009}, \emph{102} 136401.

\bibitem{Eckstein2013}
M.~Eckstein, P.~Werner,
\newblock \emph{Phys. Rev. B} \textbf{2013}, \emph{88} 075135.

\bibitem{Sentef2013}
M.~Sentef, A.~F. Kemper, B.~Moritz, J.~K. Freericks, Z.-X. Shen, T.~P.
  Devereaux,
\newblock \emph{Phys. Rev. X} \textbf{2013}, \emph{3} 041033.

\bibitem{Schueler2016}
M.~Sch\"uler, J.~Berakdar, Y.~Pavlyukh,
\newblock \emph{Phys. Rev. B} \textbf{2016}, \emph{93} 054303.

\bibitem{Mor2017}
S.~Mor, M.~Herzog, D.~Gole\ifmmode~\check{z}\else \v{z}\fi{}, P.~Werner,
  M.~Eckstein, N.~Katayama, M.~Nohara, H.~Takagi, T.~Mizokawa, C.~Monney,
  J.~St\"ahler,
\newblock \emph{Phys. Rev. Lett.} \textbf{2017}, \emph{119} 086401.

\bibitem{Perfetto2019}
E.~Perfetto, D.~Sangalli, M.~Palummo, A.~Marini, G.~Stefanucci,
\newblock \emph{Journal of Chemical Theory and Computation} \textbf{2019},
  \emph{15}, 8 4526.

\bibitem{Balaz2023}
P.~Bal\'a\ifmmode~\check{z}\else \v{z}\fi{}, M.~Zwierzycki, F.~Cosco, K.~Carva,
  P.~Maldonado, P.~M. Oppeneer,
\newblock \emph{Phys. Rev. B} \textbf{2023}, \emph{107} 174418.

\bibitem{Myohanen2012}
P.~My\"oh\"anen, R.~Tuovinen, T.~Korhonen, G.~Stefanucci, R.~van Leeuwen,
\newblock \emph{Phys. Rev. B} \textbf{2012}, \emph{85} 075105.

\bibitem{Latini2014}
S.~Latini, E.~Perfetto, A.-M. Uimonen, R.~van Leeuwen, G.~Stefanucci,
\newblock \emph{Phys. Rev. B} \textbf{2014}, \emph{89} 075306.

\bibitem{Antipov2017}
A.~E. Antipov, Q.~Dong, J.~Kleinhenz, G.~Cohen, E.~Gull,
\newblock \emph{Phys. Rev. B} \textbf{2017}, \emph{95} 085144.

\bibitem{Ridley2018}
M.~Ridley, V.~N. Singh, E.~Gull, G.~Cohen,
\newblock \emph{Phys. Rev. B} \textbf{2018}, \emph{97} 115109.

\bibitem{Ridley2019}
M.~Ridley, M.~Galperin, E.~Gull, G.~Cohen,
\newblock \emph{Phys. Rev. B} \textbf{2019}, \emph{100} 165127.

\bibitem{Basilewitsch2019}
D.~Basilewitsch, F.~Cosco, N.~L. Gullo, M.~Möttönen, T.~Ala-Nissilä, C.~P.
  Koch, S.~Maniscalco,
\newblock \emph{New Journal of Physics} \textbf{2019}, \emph{21}, 9 093054.

\bibitem{Covito2020}
F.~Covito, A.~Rubio, F.~G. Eich,
\newblock \emph{Journal of Chemical Theory and Computation} \textbf{2020},
  \emph{16}, 1 295.

\bibitem{Cohen2020}
G.~Cohen, M.~Galperin,
\newblock \emph{The Journal of Chemical Physics} \textbf{2020}, \emph{152}, 9
  090901.

\bibitem{Dutta2020}
B.~Dutta, D.~Majidi, N.~W. Talarico, N.~Lo~Gullo, H.~Courtois, C.~B.
  Winkelmann,
\newblock \emph{Phys. Rev. Lett.} \textbf{2020}, \emph{125} 237701.

\bibitem{Ridley2022}
M.~Ridley, N.~W. Talarico, D.~Karlsson, N.~L. Gullo, R.~Tuovinen,
\newblock \emph{Journal of Physics A: Mathematical and Theoretical}
  \textbf{2022}, \emph{55}, 27 273001.

\bibitem{Murakami2017}
Y.~Murakami, N.~Tsuji, M.~Eckstein, P.~Werner,
\newblock \emph{Phys. Rev. B} \textbf{2017}, \emph{96} 045125.

\bibitem{Sentef2017}
M.~A. Sentef, A.~Tokuno, A.~Georges, C.~Kollath,
\newblock \emph{Phys. Rev. Lett.} \textbf{2017}, \emph{118} 087002.

\bibitem{Li2018}
J.~Li, H.~U.~R. Strand, P.~Werner, M.~Eckstein,
\newblock \emph{Nature Communications} \textbf{2018}, \emph{9}, 1 4581.

\bibitem{Sentef2018}
M.~A. Sentef, M.~Ruggenthaler, A.~Rubio,
\newblock \emph{Science Advances} \textbf{2018}, \emph{4}, 11.

\bibitem{Topp2018}
G.~E. Topp, N.~Tancogne-Dejean, A.~F. Kemper, A.~Rubio, M.~A. Sentef,
\newblock \emph{Nature Communications} \textbf{2018}, \emph{9}, 1 4452.

\bibitem{Fausti2011}
D.~Fausti, R.~I. Tobey, N.~Dean, S.~Kaiser, A.~Dienst, M.~C. Hoffmann, S.~Pyon,
  T.~Takayama, H.~Takagi, A.~Cavalleri,
\newblock \emph{Science} \textbf{2011}, \emph{331} 189.

\bibitem{Denny2015}
S.~J. Denny, S.~R. Clark, Y.~Laplace, A.~Cavalleri, D.~Jaksch,
\newblock \emph{Phys. Rev. Lett.} \textbf{2015}, \emph{114} 137001.

\bibitem{Mitrano2016}
M.~Mitrano, A.~Cantaluppi, D.~Nicoletti, S.~Kaiser, A.~Perucchi, S.~Lupi,
  P.~Di~Pietro, D.~Pontiroli, M.~Ricc{\`o}, S.~R. Clark, D.~Jaksch,
  A.~Cavalleri,
\newblock \emph{Nature} \textbf{2016}, \emph{530} 461.

\bibitem{Werdehausen2018}
D.~Werdehausen, T.~Takayama, M.~H{\"o}ppner, G.~Albrecht, A.~W. Rost, Y.~Lu,
  D.~Manske, H.~Takagi, S.~Kaiser,
\newblock \emph{Sci. Adv.} \textbf{2018}, \emph{4} eaap8652.

\bibitem{Niedermayr2022}
A.~Niedermayr, M.~Volkov, S.~A. Sato, N.~Hartmann, Z.~Schumacher, S.~Neb,
  A.~Rubio, L.~Gallmann, U.~Keller,
\newblock \emph{Phys. Rev. X} \textbf{2022}, \emph{12} 021045.

\bibitem{Gupta2023}
R.~Gupta, F.~Cosco, R.~S. Malik, X.~Chen, S.~Saha, A.~Ghosh, T.~Pohlmann,
  J.~R.~L. Mardegan, S.~Francoual, R.~Stefanuik, J.~S\"oderstr\"om, B.~Sanyal,
  O.~Karis, P.~Svedlindh, P.~M. Oppeneer, R.~Knut,
\newblock \emph{Phys. Rev. B} \textbf{2023}, \emph{108} 064427.

\bibitem{Lindner2011}
N.~H. Lindner, G.~Refael, V.~Galitski,
\newblock \emph{Nature Physics} \textbf{2011}, \emph{7}, 6 490.

\bibitem{Wang2013}
Y.~H. Wang, H.~Steinberg, P.~Jarillo-Herrero, N.~Gedik,
\newblock \emph{Science} \textbf{2013}, \emph{342}, 6157 453.

\bibitem{Mahmood2016}
F.~Mahmood, C.-K. Chan, Z.~Alpichshev, D.~Gardner, Y.~Lee, P.~A. Lee, N.~Gedik,
\newblock \emph{Nature Physics} \textbf{2016}, \emph{12}, 4 306.

\bibitem{Huebener2017}
H.~H{\"u}bener, M.~A. Sentef, U.~De~Giovannini, A.~F. Kemper, A.~Rubio,
\newblock \emph{Nature Communications} \textbf{2017}, \emph{8}, 1 13940.

\bibitem{Kennes2019}
D.~M. Kennes, M.~Claassen, M.~A. Sentef, C.~Karrasch,
\newblock \emph{Phys. Rev. B} \textbf{2019}, \emph{100} 075115.

\bibitem{Topp2019}
G.~E. Topp, G.~Jotzu, J.~W. McIver, L.~Xian, A.~Rubio, M.~A. Sentef,
\newblock \emph{Phys. Rev. Research} \textbf{2019}, \emph{1} 023031.

\bibitem{Kaiser2014}
S.~Kaiser, C.~R. Hunt, D.~Nicoletti, W.~Hu, I.~Gierz, H.~Y. Liu, M.~Le~Tacon,
  T.~Loew, D.~Haug, B.~Keimer, A.~Cavalleri,
\newblock \emph{Phys. Rev. B} \textbf{2014}, \emph{89} 184516.

\bibitem{Hu2014}
W.~Hu, S.~Kaiser, D.~Nicoletti, C.~R. Hunt, I.~Gierz, M.~C. Hoffmann,
  M.~Le~Tacon, T.~Loew, B.~Keimer, A.~Cavalleri,
\newblock \emph{Nat. Mater.} \textbf{2014}, \emph{13} 705.

\bibitem{McIver2020}
J.~W. McIver, B.~Schulte, F.-U. Stein, T.~Matsuyama, G.~Jotzu, G.~Meier,
  A.~Cavalleri,
\newblock \emph{Nature Physics} \textbf{2020}, \emph{16}, 1 38.

\bibitem{Talarico2020}
N.~W. Talarico, S.~Maniscalco, N.~L. Gullo,
\newblock \emph{Phys. Rev. B} \textbf{2020}, \emph{101} 045103.

\bibitem{Portugal2021}
P.~Portugal, C.~Flindt, N.~Lo~Gullo,
\newblock \emph{Phys. Rev. B} \textbf{2021}, \emph{104} 205420.

\bibitem{Eckstein2011}
M.~Eckstein, P.~Werner,
\newblock \emph{Phys. Rev. B} \textbf{2011}, \emph{84} 035122.

\bibitem{Ligges2018}
M.~Ligges, I.~Avigo, D.~Gole\ifmmode~\check{z}\else \v{z}\fi{}, H.~U.~R.
  Strand, Y.~Beyazit, K.~Hanff, F.~Diekmann, L.~Stojchevska, M.~Kall\"ane,
  P.~Zhou, K.~Rossnagel, M.~Eckstein, P.~Werner, U.~Bovensiepen,
\newblock \emph{Phys. Rev. Lett.} \textbf{2018}, \emph{120} 166401.

\bibitem{Peronaci2018}
F.~Peronaci, M.~Schir\'o, O.~Parcollet,
\newblock \emph{Phys. Rev. Lett.} \textbf{2018}, \emph{120} 197601.

\bibitem{Johnson2016}
T.~H. Johnson, F.~Cosco, M.~T. Mitchison, D.~Jaksch, S.~R. Clark,
\newblock \emph{Phys. Rev. A} \textbf{2016}, \emph{93} 053619.

\bibitem{LoGullo2016}
N.~Lo~Gullo, L.~Dell'Anna,
\newblock \emph{Phys. Rev. B} \textbf{2016}, \emph{94} 184308.

\bibitem{Cosco2017}
F.~Cosco, M.~Borrelli, P.~Silvi, S.~Maniscalco, G.~De~Chiara,
\newblock \emph{Phys. Rev. A} \textbf{2017}, \emph{95} 063615.

\bibitem{Cosco2017Pr}
F.~Cosco, M.~Borrelli, F.~Plastina, S.~Maniscalco,
\newblock \emph{Phys. Rev. A} \textbf{2017}, \emph{95} 053620.

\bibitem{Cosco2018Me}
F.~Cosco, S.~Maniscalco,
\newblock \emph{Phys. Rev. A} \textbf{2018}, \emph{98} 053608.

\bibitem{Cosco2018AU}
F.~Cosco, M.~Borrelli, E.-M. Laine, S.~Pascazio, A.~Scardicchio, S.~Maniscalco,
\newblock \emph{New Journal of Physics} \textbf{2018}, \emph{20}, 7 073041.

\bibitem{Settino2020AU}
J.~Settino, N.~W. Talarico, F.~Cosco, F.~Plastina, S.~Maniscalco, N.~Lo~Gullo,
\newblock \emph{Phys. Rev. B} \textbf{2020}, \emph{101} 144303.

\bibitem{Cassette2015}
E.~Cassette, R.~D. Pensack, B.~Mahler, G.~D. Scholes,
\newblock \emph{Nature Communications} \textbf{2015}, \emph{6}, 1 6086.

\bibitem{Kemper2018}
A.~F. Kemper, O.~Abdurazakov, J.~K. Freericks,
\newblock \emph{Phys. Rev. X} \textbf{2018}, \emph{8} 041009.

\bibitem{Aspelmeyer2014}
M.~Aspelmeyer, T.~J. Kippenberg, F.~Marquardt,
\newblock \emph{Rev. Mod. Phys.} \textbf{2014}, \emph{86} 1391.

\bibitem{Cosco2021}
F.~Cosco, J.~S. Pedernales, M.~B. Plenio,
\newblock \emph{Phys. Rev. A} \textbf{2021}, \emph{103} L061501.

\bibitem{Rusconi2022}
C.~C. Rusconi, M.~Perdriat, G.~H\'etet, O.~Romero-Isart, B.~A. Stickler,
\newblock \emph{Phys. Rev. Lett.} \textbf{2022}, \emph{129} 093605.

\bibitem{Ballestero2023}
C.~Gonzalez-Ballestero, J.~Zieli\ifmmode~\acute{n}\else \'{n}\fi{}ska,
  M.~Rossi, A.~Militaru, M.~Frimmer, L.~Novotny, P.~Maurer, O.~Romero-Isart,
\newblock \emph{PRX Quantum} \textbf{2023}, \emph{4} 030331.

\bibitem{Stefanucci2013}
G.~Stefanucci, R.~van Leeuwen,
\newblock \emph{Nonequilibrium Many-Body Theory of Quantum Systems: A Modern
  Introduction},
\newblock Cambridge University Press, \textbf{2013}.

\bibitem{Aoki2014}
H.~Aoki, N.~Tsuji, M.~Eckstein, M.~Kollar, T.~Oka, P.~Werner,
\newblock \emph{Rev. Mod. Phys.} \textbf{2014}, \emph{86} 779.

\bibitem{Kurth2005}
S.~Kurth, G.~Stefanucci, C.-O. Almbladh, A.~Rubio, E.~K.~U. Gross,
\newblock \emph{Phys. Rev. B} \textbf{2005}, \emph{72} 035308.

\bibitem{Lipavsky1986}
P.~Lipavsk\'y, V.~\ifmmode \check{S}\else \v{S}\fi{}pi\ifmmode~\check{c}\else
  \v{c}\fi{}ka, B.~Velick\'y,
\newblock \emph{Phys. Rev. B} \textbf{1986}, \emph{34} 6933.

\bibitem{Spicka2005}
V.~\ifmmode\check{S}\else\v{S}\fi{}pi\ifmmode\check{c}\else\v{c}\fi{}ka,
  B.~Velick\'y, A.~Kalvov\'a,
\newblock \emph{Physica E: Low-dimensional Systems and Nanostructures}
  \textbf{2005}, \emph{29}, 1 154 .

\bibitem{Balzer2013gkba}
K.~Balzer, S.~Hermanns, M.~Bonitz,
\newblock \emph{Journal of Physics: Conference Series} \textbf{2013},
  \emph{427} 012006.

\bibitem{Hermanns2012}
S.~Hermanns, K.~Balzer, M.~Bonitz,
\newblock \emph{Physica Scripta} \textbf{2012}, \emph{T151} 014036.

\bibitem{Hermanns2013}
S.~Hermanns, K.~Balzer, M.~Bonitz,
\newblock \emph{Journal of Physics: Conference Series} \textbf{2013},
  \emph{427} 012008.

\bibitem{Haug1996}
H.~Haug, L.~B{\'a}nyai,
\newblock \emph{Solid State Communications} \textbf{1996}, \emph{100}, 5 303 .

\bibitem{Bonitz1996}
M.~Bonitz, D.~Kremp, D.~C. Scott, R.~Binder, W.~D. Kraeft, H.~S. Köhler,
\newblock \emph{Journal of Physics: Condensed Matter} \textbf{1996}, \emph{8},
  33 6057.

\bibitem{Bonitz1999}
M.~Bonitz, D.~Semkat, H.~Haug,
\newblock \emph{The European Physical Journal B - Condensed Matter and Complex
  Systems} \textbf{1999}, \emph{9}, 2 309.

\bibitem{Kwong1998}
N.~H. Kwong, M.~Bonitz, R.~Binder, H.~S. Köhler,
\newblock \emph{physica status solidi (b)} \textbf{1998}, \emph{206}, 1 197.

\bibitem{Pal2009}
{Pal, G.}, {Pavlyukh, Y.}, {Schneider, H. C.}, {H\"ubner, W.},
\newblock \emph{Eur. Phys. J. B} \textbf{2009}, \emph{70}, 4 483.

\bibitem{Hopjan2018}
M.~Hopjan, G.~Stefanucci, E.~Perfetto, C.~Verdozzi,
\newblock \emph{Phys. Rev. B} \textbf{2018}, \emph{98} 041405.

\bibitem{Perfetto2018CHEERS}
E.~Perfetto, G.~Stefanucci,
\newblock \emph{Journal of Physics: Condensed Matter} \textbf{2018}, \emph{30},
  46 465901.

\bibitem{Karlsson2018}
D.~Karlsson, R.~van Leeuwen, E.~Perfetto, G.~Stefanucci,
\newblock \emph{Phys. Rev. B} \textbf{2018}, \emph{98} 115148.

\bibitem{Hopjan2019}
M.~Hopjan, C.~Verdozzi,
\newblock \emph{Eur. Phys. J. Spec. Top.} \textbf{2019}, \emph{227}, 15 1939.

\bibitem{Bonitz2019}
M.~Bonitz, K.~Balzer, N.~Schl{\"u}nzen, M.~R. Rasmussen, J.-P. Joost,
\newblock \emph{Phys. Status Solidi B} \textbf{2019}, \emph{256}, 7 1800490.

\bibitem{Schlunzen2017}
N.~Schl\"unzen, J.-P. Joost, F.~Heidrich-Meisner, M.~Bonitz,
\newblock \emph{Phys. Rev. B} \textbf{2017}, \emph{95} 165139.

\bibitem{Schlunzen2019}
N.~Schlünzen, S.~Hermanns, M.~Scharnke, M.~Bonitz,
\newblock \emph{Journal of Physics: Condensed Matter} \textbf{2019}, \emph{32},
  10 103001.

\bibitem{Schlunzen2020}
N.~Schl\"unzen, J.-P. Joost, M.~Bonitz,
\newblock \emph{Phys. Rev. Lett.} \textbf{2020}, \emph{124} 076601.

\bibitem{Joost2020}
J.-P. Joost, N.~Schl\"unzen, M.~Bonitz,
\newblock \emph{Phys. Rev. B} \textbf{2020}, \emph{101} 245101.

\bibitem{Karlsson2021}
D.~Karlsson, R.~van Leeuwen, Y.~Pavlyukh, E.~Perfetto, G.~Stefanucci,
\newblock \emph{Phys. Rev. Lett.} \textbf{2021}, \emph{127} 036402.

\bibitem{Tuovinen2023}
R.~Tuovinen, Y.~Pavlyukh, E.~Perfetto, G.~Stefanucci,
\newblock \emph{Phys. Rev. Lett.} \textbf{2023}, \emph{130} 246301.

\bibitem{Pyykkonen2021}
V.~A.~J. Pyykk\"onen, S.~Peotta, P.~Fabritius, J.~Mohan, T.~Esslinger,
  P.~T\"orm\"a,
\newblock \emph{Phys. Rev. B} \textbf{2021}, \emph{103} 144519.

\bibitem{Pyykkonen2023}
V.~A.~J. Pyykk\"onen, S.~Peotta, P.~T\"orm\"a,
\newblock \emph{Phys. Rev. Lett.} \textbf{2023}, \emph{130} 216003.

\bibitem{Talarico2019}
N.~W. Talarico, S.~Maniscalco, N.~Lo~Gullo,
\newblock \emph{physica status solidi (b)} \textbf{2019}, \emph{256}, 7
  1800501.

\bibitem{Hermanns2014}
S.~Hermanns, N.~Schl\"unzen, M.~Bonitz,
\newblock \emph{Phys. Rev. B} \textbf{2014}, \emph{90} 125111.

\bibitem{Bostrom2018}
E.~V. Bostr{\"o}m, A.~Mikkelsen, C.~Verdozzi, E.~Perfetto, G.~Stefanucci,
\newblock \emph{Nano Letters} \textbf{2018}, \emph{18}, 2 785, pMID: 29266952.

\bibitem{Perfetto2018}
E.~Perfetto, D.~Sangalli, A.~Marini, G.~Stefanucci,
\newblock \emph{The Journal of Physical Chemistry Letters} \textbf{2018},
  \emph{9}, 6 1353, pMID: 29494772.

\bibitem{Tuovinen2019pssb}
R.~Tuovinen, D.~Gole\ifmmode~\check{z}\else \v{z}\fi{}, M.~Sch{\"u}ler,
  P.~Werner, M.~Eckstein, M.~A. Sentef,
\newblock \emph{Phys. Status Solidi B} \textbf{2019}, \emph{256} 1800469.

\bibitem{Schueler2019}
M.~Sch\"uler, J.~C. Budich, P.~Werner,
\newblock \emph{Phys. Rev. B} \textbf{2019}, \emph{100} 041101.

\bibitem{Perfetto2020}
E.~Perfetto, A.~Trabattoni, F.~Calegari, M.~Nisoli, A.~Marini, G.~Stefanucci,
\newblock \emph{The Journal of Physical Chemistry Letters} \textbf{2020},
  \emph{11}, 3 891, pMID: 31944766.

\bibitem{Murakami2020}
Y.~Murakami, M.~Sch\"uler, S.~Takayoshi, P.~Werner,
\newblock \emph{Phys. Rev. B} \textbf{2020}, \emph{101} 035203.

\bibitem{Schueler2020}
M.~Sch\"uler, U.~De~Giovannini, H.~H\"ubener, A.~Rubio, M.~A. Sentef, T.~P.
  Devereaux, P.~Werner,
\newblock \emph{Phys. Rev. X} \textbf{2020}, \emph{10} 041013.

\bibitem{Bostrom2019}
E.~V. Bostr{\"o}m, C.~Verdozzi,
\newblock \emph{physica status solidi (b)} \textbf{2019}, \emph{256}, 7
  1800590.

\bibitem{Tuovinen2020}
R.~Tuovinen, D.~Gole\ifmmode~\check{z}\else \v{z}\fi{}, M.~Eckstein, M.~A.
  Sentef,
\newblock \emph{Phys. Rev. B} \textbf{2020}, \emph{102} 115157.

\bibitem{Tuovinen2021JCP}
R.~Tuovinen, R.~van Leeuwen, E.~Perfetto, G.~Stefanucci,
\newblock \emph{The Journal of Chemical Physics} \textbf{2021}, \emph{154}, 9
  094104.

\bibitem{Tuovinen2021NJP}
R.~Tuovinen,
\newblock \emph{New Journal of Physics} \textbf{2021}, \emph{23}, 8 083024.

\bibitem{Kalvova2017}
A.~Kalvová, B.~Velický, V.~Špička,
\newblock \emph{Journal of Superconductivity and Novel Magnetism}
  \textbf{2017}, \emph{30} 807.

\bibitem{Kalvova2019}
E.~Kalvová, B.~R. Velick, V.~Spi, A.~Kalvová, V.~Spi, B.~Velick,
\newblock \emph{physica status solidi (b)} \textbf{2019}, \emph{256} 1800594.

\bibitem{Kalvova2023}
A.~Kalvová, V.~Špička, B.~Velický, P.~Lipavský,
\newblock \emph{Europhysics Letters} \textbf{2023}, \emph{141} 16002.

\bibitem{Tuovinen2019-2B}
R.~Tuovinen, F.~Covito, M.~A. Sentef,
\newblock \emph{J. Chem. Phys.} \textbf{2019}, \emph{151}, 17 174110.

\bibitem{Pavlyukh2022}
Y.~Pavlyukh, E.~Perfetto, D.~Karlsson, R.~van Leeuwen, G.~Stefanucci,
\newblock \emph{Phys. Rev. B} \textbf{2022}, \emph{105} 125135.

\bibitem{Bruus_Flensberg}
H.~Bruus, K.~Flensberg,
\newblock \emph{Many-body quantum theory in condensed matter physics - an
  introduction},
\newblock Oxford University Press, United States, \textbf{2004}.

\bibitem{Meir1992}
Y.~Meir, N.~S. Wingreen,
\newblock \emph{Phys. Rev. Lett.} \textbf{1992}, \emph{68} 2512.

\bibitem{Jauho1994}
A.-P. Jauho, N.~S. Wingreen, Y.~Meir,
\newblock \emph{Phys. Rev. B} \textbf{1994}, \emph{50} 5528.

\bibitem{Tuovinen2013}
R.~Tuovinen, R.~van Leeuwen, E.~Perfetto, G.~Stefanucci,
\newblock \emph{Journal of Physics: Conference Series} \textbf{2013},
  \emph{427} 012014.

\bibitem{Landauer1957}
R.~{Landauer},
\newblock \emph{IBM Journal of Research and Development} \textbf{1957},
  \emph{1}, 3 223.

\bibitem{Buttiker1986}
M.~B\"uttiker,
\newblock \emph{Phys. Rev. Lett.} \textbf{1986}, \emph{57} 1761.

\bibitem{Jauho_Haug}
H.~Haug, A.-P. Jauho,
\newblock \emph{Quantum kinetics in transport and optics of semiconductors;
  2nd, substantially rev. ed.}, volume 123 of \emph{Springer series in
  solid-state Sciences},
\newblock Springer, Berlin, \textbf{2008}.

\bibitem{Mielke1991}
A.~Mielke,
\newblock \emph{Journal of Physics A: Mathematical and General} \textbf{1991},
  \emph{24}, 2 L73.

\bibitem{Tasaki1992}
H.~Tasaki,
\newblock \emph{Physical Review Letters} \textbf{1992}, \emph{69}, 10 1608.

\bibitem{Mielke1993}
A.~Mielke, H.~Tasaki,
\newblock \emph{Communications in Mathematical Physics} \textbf{1993},
  \emph{158}, 2 341.

\bibitem{Rubio2008}
K.~S. Thygesen, A.~Rubio,
\newblock \emph{Phys. Rev. B} \textbf{2008}, \emph{77} 115333.

\bibitem{Bordia2017}
P.~Bordia, H.~L\"uschen, S.~Scherg, S.~Gopalakrishnan, M.~Knap, U.~Schneider,
  I.~Bloch,
\newblock \emph{Phys. Rev. X} \textbf{2017}, \emph{7} 041047.

\bibitem{Joost2022}
J.-P. Joost, N.~Schl\"unzen, H.~Ohldag, M.~Bonitz, F.~Lackner,
  I.~B\ifmmode~\check{r}\else \v{r}\fi{}ezinov\'a,
\newblock \emph{Phys. Rev. B} \textbf{2022}, \emph{105} 165155.

\bibitem{Thygesen2008}
K.~S. Thygesen,
\newblock \emph{Phys. Rev. Lett.} \textbf{2008}, \emph{100} 166804.

\bibitem{Stefanucci2010}
G.~Stefanucci, E.~Perfetto, M.~Cini,
\newblock \emph{Phys. Rev. B} \textbf{2010}, \emph{81} 115446.

\bibitem{Jiang2011}
L.~Jiang, T.~Kitagawa, J.~Alicea, A.~R. Akhmerov, D.~Pekker, G.~Refael, J.~I.
  Cirac, E.~Demler, M.~D. Lukin, P.~Zoller,
\newblock \emph{Phys. Rev. Lett.} \textbf{2011}, \emph{106} 220402.

\bibitem{Weston2015}
J.~Weston, B.~Gaury, X.~Waintal,
\newblock \emph{Phys. Rev. B} \textbf{2015}, \emph{92} 020513.

\bibitem{Francica2016}
G.~Francica, T.~J.~G. Apollaro, N.~Lo~Gullo, F.~Plastina,
\newblock \emph{Phys. Rev. B} \textbf{2016}, \emph{94} 245103.

\bibitem{Tuovinen2016PNGF}
R.~Tuovinen, R.~van Leeuwen, E.~Perfetto, G.~Stefanucci,
\newblock \emph{Journal of Physics: Conference Series} \textbf{2016},
  \emph{696} 012016.

\bibitem{Thakurathi2017}
M.~Thakurathi, D.~Loss, J.~Klinovaja,
\newblock \emph{Phys. Rev. B} \textbf{2017}, \emph{95} 155407.

\bibitem{Dehghani2017}
H.~Dehghani, A.~Mitra,
\newblock \emph{Phys. Rev. B} \textbf{2017}, \emph{96} 195110.

\bibitem{Claassen2019}
M.~Claassen, D.~M. Kennes, M.~Zingl, M.~A. Sentef, A.~Rubio,
\newblock \emph{Nature Physics} \textbf{2019}, \emph{15}, 8 766.

\bibitem{Tuovinen2019NJP}
R.~Tuovinen, E.~Perfetto, R.~van Leeuwen, G.~Stefanucci, M.~A. Sentef,
\newblock \emph{New J. Phys.} \textbf{2019}, \emph{21}, 10 103038.

\end{thebibliography}
\end{document}